\documentclass[a4paper,12pt]{article}
\usepackage[latin1]{inputenc}
\usepackage[T1]{fontenc}
\usepackage{amssymb}
\usepackage{amsmath}
\usepackage{graphicx}
\usepackage{pstricks}

\numberwithin{equation}{section}

\newcommand{\RR}{\mathbb{R}}

\newcommand{\be}{\begin{equation}}
\newcommand{\ee}{\end{equation}}
\newcommand{\bea}{\begin{eqnarray}}
\newcommand{\eea}{\end{eqnarray}}
\newcommand{\ud}{\mathrm{d}}
\newcommand{\G}{\left}
\newcommand{\D}{\right}
\newcommand{\p}{\partial}
\newcommand{\bp}{\overline{\partial}}
\newcommand{\bz}{{\bar{z}}}

\newcommand{\cA}{\mathcal{A}}
\newcommand{\cV}{\mathcal{V}}
\newcommand{\cG}{\mathcal{G}}
\newcommand{\cS}{\mathcal{S}}
\newcommand{\cT}{\mathcal{T}}

\newcommand{\tG}{\widetilde{G}}

\newcommand{\w}{\wedge}

\newcommand{\ie}{\emph{i.e. }}

\newcommand{\ath}{\theta}
\newcommand{\bth}{{\bar\theta}}

\newcommand{\bD}{\overline{D}}
\newcommand{\bG}{\overline{G}}
\newcommand{\bX}{\overline{X}}
\newcommand{\bY}{\overline{Y}}
\newcommand{\cY}{\mathcal{Y}}
\newcommand{\cX}{\mathcal{X}}
\newcommand{\cD}{\mathcal{D}}
\newcommand{\bcD}{\overline{\mathcal{D}}}
\newcommand{\tcV}{\widetilde{\mathcal{V}}}
\newcommand{\tcG}{\widetilde{\mathcal{G}}}
\newcommand{\tcX}{\widetilde{\mathcal{X}}}
\newcommand{\tcY}{\widetilde{\mathcal{Y}}}
\newcommand{\btG}{\widetilde{\overline{G}}}
\newcommand{\tfG}{\widetilde{\mathfrak{G}}}
\newcommand{\tfg}{\widetilde{\mathfrak{g}}}
\newcommand{\tfH}{\widetilde{\mathfrak{H}}}
\newcommand{\tfh}{\widetilde{\mathfrak{h}}}

\title{Superconformal Selfdual $\sigma$-Models}

\author{Louis Paulot}

\date{}

\begin{document}

\begin{flushright}
ULB-TH/04-10\\
April 2004\\
\vspace*{1cm}
\end{flushright}

\begin{center}
\begin{Large}
\textbf{Superconformal Selfdual $\sigma$-Models}
\end{Large}

\vspace{7mm}
{\bf Louis Paulot}

\vspace{5mm}
Physique théorique et mathématique\\
and International Solvay Institutes\\
Université libre de Bruxelles\\
C.P.~231, B-1050 Bruxelles, Belgium
\vspace{3mm}

{\ttfamily lpaulot@ulb.ac.be}
\end{center}

\vspace{3mm}
\hrule
\begin{abstract} 
A range of bosonic models can be expressed as (sometimes generalized)
$\sigma$-models, with equations of motion coming from a selfduality
constraint. We show that in $D=2$, this is easily extended to 
supersymmetric cases, in a superspace approach. In particular, we find
that the configurations of fields of a superconformal
$\mathfrak{G}/\mathfrak{H}$ coset models which
satisfy some selfduality constraint are automatically solutions to the
equations of motion of the model. Finally, we show that symmetric space
$\sigma$-models can be seen as infinite-dimensional $\tfG/\tfH$ models 
constrained by a selfduality equation, with $\tfG$ the loop extension 
of $\mathfrak{G}$ and $\tfH$ a maximal subgroup. It ensures that these 
models have a hidden global $\tfG$ symmetry together with a local $\tfH$ 
gauge symmetry.
\end{abstract}
\hrule

\section{Introduction}

Several (classical) bosonic models have been expressed as generalized 
$\sigma$-models constrained by a selfduality equation. In this formalism, 
the set of fields is doubled and equations of motion are given by Bianchi
identities of dual fields. This class of models contains the bosonic matter 
sector of supergravities and especially eleven-dimensional supergravity and
reductions of these on tori, up to $D=3$
\cite{cjlp2,hjp1,hjp2,west1,s-w1,s-w2}, as well as all models 
obtained by oxidation of $D=3$ principal $\sigma$-models for all simple 
groups \cite{cjlp3,dhjp,ehtw}, and is closely related to the so-called
"$E_{11}$ conjecture" \cite{west1,ju2}.

The inclusion of the fermionic and gravity sectors of such models into this
formalism has not yet been achieved, although there are some results on
$D=2$ gravity models coming from $D=3$ coset models
\cite{brei-mai,ju1,ni2,ju-ni,ber-ju}, sometimes including fermions in the
case of $N=16$ supergravity \cite{ni1,ni-wa,ni3,ni-sa}. These $D=2$ models
can indeed be seen as integrable systems, which opens a path to quantization
\cite{kn,ks1,ks2,kns}. One can also find attempts of including fermions
in the doubled formalism for eleven-dimensionnal and type IIA supergravities
in \cite{nur1,nur2}.

In a program of including fermions, gravity and supersymmetry into such a
formalism for generalized $\sigma$-models, especially in higher dimension,
and revealing hidden symmetries of theories, we begin here by studying $D=2$
models without (super)gravity. We extend selfduality of $\sigma$-models to
superconformal cases, by defining a Hodge star on superspace that we use
together with a pseudo-involution. Special cases
have already been studied in the context of solitonic and instantonic
solutions of some supersymmetric $\sigma$-models, starting from
\cite{dv-f,w,fgs}. Here we develop a general theory where the target-space
can be any Lie group or symmetric space.

After recalling basic facts on the selfdual formalism in section
\ref{sigma}, we introduce its superconformal version (section
\ref{super-rigid}); we study then solutions to a selfduality constraint in a
superconformal symmetric space $\sigma$-model in section \ref{self-sym}, and
we show that selfdual sets of fields are special solutions to the usual
equations of motion.  As a simple example, we treat very explicitely the
case of $SL(2)/SO(2)$.

Our main result is in section \ref{loop}: we give a formulation of such $\mathfrak{G}/\mathfrak{H}$
$\sigma$-models as selfdual infinite $\tfG/\tfH$ $\sigma$-models, where
$\tfG$ is the loop extension of the group $\mathfrak{G}$, and $\tfH$ a
maximal subgroup containing $\mathfrak{H}$. Physical fields with their
Bianchi identities and equations of motion appear through the Lax pair
associated to the model. The hidden group of symmetries of such models,
including non-local ones, have been discussed for some time
\cite{hruby,cz,cgw,dolan,ei-fo,wu,jsa,schwarz} with different conclusions.
With our construction, it is manifest that in addition to (super)conformal
symmetry on the worldsheet, there is a global symmetry $\tfG$ and a local
gauge symmetry $\tfH$. It has been checked in some cases that such
symmetries survive quantization (\cite{eky} and references therein.)

\section{Selfdual $\sigma$-models}
\label{sigma}

\subsection{$\sigma$-models}

A $\sigma$-model is described by a map $\phi$ from spacetime $\Sigma$
to a target manifold $\mathcal{M}$. Let $n$ be the dimension of the target
space. 

Here we specify to the case where $\mathcal{M}$ is a simply connected
Lie group $\mathfrak{G}$. Let
$\{T_i\}$ be a basis of generators of the tangent Lie algebra
$\mathfrak{g}$. Elements of $\mathfrak{G}$ can be parametrized locally, in
the vicinity of $\cV_0$, as
\be
\cV = \exp \G( \sum_i \phi^i T_i \D) \, \cV_0
\ee
with $n$ scalars $\phi^i$, where $n$ is the dimension of $\mathfrak{G}$.

On $\mathcal{M} = \mathfrak{G}$, we have the $\mathfrak{g}$-valued
Maurer-Cartan form
\be
\sigma = \p_i \cV \, \cV^{-1} \ud\phi^i\rlap{\ ,}
\label{vielbein}
\ee
where $\p_i = \frac{\p}{\p\phi^i}$.
We can compute it with help of the Baker-Campbell-Hausdorff formula
\be
\p_i \cV \, \cV^{-1} =  \p_i e^\phi \, e^{-\phi}
= \p_i \phi + \frac{1}{2!} [ \phi, \p_i \phi]
+ \frac{1}{3!} [\phi, [\phi,  \p_i \phi]] + \ldots
\ee
(where $\phi = \sum \phi^j T_j$).

We define the field strength associated to $\phi$ as the pullback on
$\Sigma$ of the Maurer-Cartan form on $\mathfrak{G}$.
\be
\cG = \ud \cV \, \cV^{-1} = \p_\mu\phi^i \, \p_i\cV \, \cV^{-1} \ud x^\mu
\rlap{\ ,}
\ee
where $\p_i$ is the target-space derivative and $\p_\mu$ the spacetime
derivative, with respect to $x^\mu$.

This field strength is invariant under action of $\mathfrak{G}$.
Indeed, if one acts on
$\cV$:
\be
\cV \longrightarrow \cV \Lambda
\ee
with elements
\be
\Lambda = e^{\G( \lambda^i T_i\D)}
\ee
defined with closed scalars $\lambda^i$ ($\ud \lambda^i=0$), $\cG$ becomes
\be
\cG \longrightarrow \ud\cV \Lambda \Lambda^{-1}\cV^{-1} + \cV \ud \Lambda
\Lambda^{-1}\cV^{-1} \rlap{\ ,}
\ee
where the second term vanishes because of $\ud \lambda = 0$.

The $\sigma$-model structure we have recalled in this part is valid for any
number of spacetime dimensions. We restrict now to two dimensions for the
rest of the paper. Moreover, we will work in Euclidean signature.
We will indicate how to transpose to Lorentzian signature.

In Euclidean signature\footnote{In a Lorentzian setting, we would use
lightcone coordinates $x^+$ and $x^-$.}, it will be convenient to use
complex notations and to write the field strength as
\bea
\cG &=& G_z \ud z + G_\bz \ud \bz \nonumber\\
&=& \p_z\phi^i \, \p_i\cV \, \cV^{-1} \ud z +
\p_\bz \phi^i \, \p_i\cV \, \cV^{-1} \ud \bz
\rlap{\ .}
\eea
In order to make the contact with the supersymmetric generalization easier,
we also think of this 1-form field as a vector in the cotangent bundle:
\be
\cG= \G( \begin{array}{c} G_z\\ G_\bz \end{array} \D)
\rlap{\ .}
\ee

\subsection{Selfduality}

We consider the group element $\cV$ as a
\emph{doubled} set of fields: equations of motion are given by a
selfduality equation.

Namely, let $\cS$ be a pseudo involution of $\mathfrak{g}$, exchanging
$T_a$'s, such that
\be
*^2 \cS^2 = 1 \rlap{\ .}
\ee
If we choose a euclidean signature, we have $*^2 = -1$ for 1-forms, 
and therefore $\cS^2=-1$. For a Lorentzian signature, it would be $*^2 = 1$
for 1-forms, and $\cS^2=1$.

Thus we can write a selfduality equation
\be
* \cS \cG = \cG \rlap{\ ,}
\label{sigma-dual}
\ee
where $*$ acts on the differentials $\cG^a$ and $\cS$ on generators $T_a$.

Following from the definition of $\cG$, we have the Bianchi identity 
\be
\ud \cG - \cG \w \cG = 0 \rlap{\ .} 
\ee

If we use the selfduality relation (\ref{sigma-dual}) to reduce by one half
the number of fields, one half of the set of Bianchi identities becomes the
equations of motion of the physical fields. Indeed we have
\be
\ud \!* \! \cS \cG - \cG \w \cG = 0 \rlap{\ .}
\ee

A trivial example is the free scalar $\varphi$. We take an abelian algebra of
dimension two, with generators $h_1$ and $h_2$. Let 
\be
\phi = \phi_1 h_1 + \phi_2 h_2 \rlap{\ ;}
\ee
the curvature is
\be
\cG = \ud \phi_1 \, h_1 + \ud \phi_2 \, h_2 \rlap{\ .}
\ee
With $\cS$ exchanging $h_1$ and $h_2$, 
\bea
\cS h_1 &=& h_2 \nonumber\\
\cS h_2 &=& -h_1 \rlap{\ ,}
\eea
the second Bianchi identity
\be
\ud \ud \phi_2 = 0
\ee
becomes the equation of motion of $\phi_1$:
\be
\ud \! * \! \ud \phi_1 = 0 \rlap{\ .}
\ee

In complex notation, in the conformal gauge, the Hodge
star $*$ has a very simple expression, which will be convenient for the
supersymmetric generalization:
\be
*(G_z \ud z + G_{\bar z} \ud \bar z) = -i G_z \ud z +i G_{\bar z} \ud \bar z
\rlap{\ .}
\ee

We remark that the selfduality equation is conformally invariant, so
that the model has (spacetime) conformal symmetry, with
$G_z$ of weight $(1,0)$ and $G_{\bar z}$ of weight $(0,1)$. Indeed, we have
\be
G_{z'} = \p_{z'} \cV \,\cV^{-1} = (\p_{z'}z)  \p_{z} \cV \,\cV^{-1}
= (\p_{z'}z) G_z \rlap{\ ,}
\ee
with the analougous for $G_{\bar z}$.

\section{Supersymmetric version}
\label{super-rigid}

\subsection{$N=(1,1)$ extension}
\label{sigma-11}

We work now in $D=2$ $N=(1,1)$ superspace: 
we have two odd variables $\ath$ and $\bth$. Supersymmetry generators are
\bea
Q&=&\p_\ath - \ath \p_z \nonumber\\
\overline{Q}&=&\p_\bth - \bth \p_\bz
\eea
and covariant derivatives
\bea
D&=&\p_\ath + \ath \p_z \nonumber\\
\bD &=&\p_\bth + \bth \p_\bz \rlap{\ .}
\eea

We construct a super-$\sigma$-model with a scalar superfield $\cV$ with
values in $\mathfrak{G}$
\be
\cV = e^\Phi \, \cV_0 \rlap{\ ,}
\ee
where
\be
\Phi = \sum_i \Phi^i T_i
\ee
is a scalar superfield with values in $\mathfrak{g}$.

The field strength is defined as:
\be
\cG= \G( \begin{array}{c}G\\ \bG \end{array} \D) 
= \G( \begin{array}{c} D\cV \, \cV^{-1}\\  \bD \cV \, \cV^{-1}\end{array} \D)
\rlap{\ .}
\ee

$G$ and $\bG$ can be computed from $\Phi$ with help of the
Baker-Campbell-Hausdorff formula:
\be
G = D e^\Phi \, e^{-\Phi}
= D \Phi + \frac{1}{2!} [ \Phi, D \Phi]
+ \frac{1}{3!} [\Phi, [\Phi, D \Phi]] + \ldots
\ee
and the analogous for $\bG$ with $\bD$.

A Bianchi identity follows from the definition of $\cG$:
\be
\bD G + D \bG = \G[ G,\bG \D] \rlap{\ .}
\ee
As $G$ and $\bG$ are odd superfields, the bracket is in fact an
anticommutator.

We define a "Hodge star" by
\be
* \cG = * \G( \begin{array}{c}G\\ \bG \end{array} \D)
=\G( \begin{array}{r}-i \, G\\ i \, \bG \end{array} \D)
\rlap{\ .}
\ee

Together with the pseudo-involution $\cS$ defined as in the purely bosonic
case, we can write a selfduality equation
\be
* \cS \cG = \cG \rlap{\ ,}
\ee
which reduces the number of physical fields by one half and turns one half of
Bianchi identities into equations of motion:
\be
-i \bD (\cS G) +i D (\cS \bG) = \G[ G,\bG \D] \rlap{\ .}
\ee

This model is invariant under a global $\mathfrak{G}$. We act on $\cV$ by
$\Lambda = e^\lambda$,
\be
\cV \longrightarrow \cV \Lambda \rlap{\ ,}
\ee
with $\lambda$ a $\mathfrak{g}$-valued superfield.
$\cG$ is invariant provided
\bea
D\lambda &=&0 \nonumber\\
\bD \lambda &=&0 \rlap{\ ,}
\eea
\ie if and only if $\lambda$ is a constant field: $\Lambda$ is an element of
$\mathfrak{G}$.

As all operators we use are covariant with respect to supersymmetry
generators $Q$ and $\overline{Q}$, we get a manifestly supersymmetric
theory. Moreover, it is clear that the selfduality equation is
superconformally invariant, with $G$ and $\bG$ of respective
weights $(\frac{1}{2},0)$ and $(0,\frac{1}{2})$. Thus the selfdual
$\sigma$-model we have defined have superconformal symmetry in addition to
its $\mathfrak{G}$ target-space symmetry.

We want to show now that, if we truncate the superfield
\be
\Phi = \phi + \ath \psi + \bth \widetilde{\psi} + \ath\bth F
\ee
to its first component
\be
\phi = \sum_i \phi^i T_i\rlap{\ ,}
\ee
we recover precisely the bosonic selfdual sigma model defined in section
\ref{sigma}.

Indeed, we get
\be
G = D e^\phi \, e^{-\phi}
= \ath \p_z \phi + \frac{1}{2!} [ \phi, \ath \p_z \phi]
+ \frac{1}{3!} [\phi, [\phi, \ath \p_z \phi]] + \ldots = \ath \, G_z^{(0)}
\ee
and
\be
\bG = \bD e^\phi \, e^{-\phi}
= \bth \p_\bz \phi + \frac{1}{2!} [ \phi, \bth \p_\bz \phi]
+ \frac{1}{3!} [\phi, [\phi, \bth \p_\bz \phi]] + \ldots = \bth \, G_{\bar
z}^{(0)}
\rlap{\ ,}
\ee
where we denote by $\cG^{(0)} = G_z^{(0)} \ud z +
G_{\bar z}^{(0)} \ud {\bar z}$ the bosonic field strength derived from
$\phi$: 
\be
\cG^{(0)} = \ud \! \G( e^\phi\D) e^{-\phi}
\rlap{\ .}
\ee
The Bianchi identity becomes
\be
-\ath\bth\p_\bz G_z^{(0)} + \ath\bth\p_z G_{\bar z}^{(0)} 
= \ath\bth \G[ G_z^{(0)},  G_{\bar z}^{(0)} \D]
\rlap{\ ,}
\ee
which can be written for the bosonic field strength $\cG^{(0)}$ as
\be
\ud \cG^{(0)} - \cG^{(0)} \w \cG^{(0)} = 0 \rlap{\ .}
\ee

The Hodge star becomes
\be
* \G( \begin{array}{c}\ath G_z^{(0)} \\ \bth  G_{\bar z}^{(0)} \end{array} \D)
=\G( \begin{array}{r} -i\ath G_z^{(0)} \\ i \bth  G_{\bar z}^{(0)} \end{array} \D)
\rlap{\ :}
\ee
it reduces to the usual Hodge star in conformal gauge for $\cG^{(0)}$.

As a consequence, the self duality equation
\be
* \cS \cG = \cG
\ee
gives for the truncated superfield $\Phi = \phi$ 
the same equations of motion as in the purely bosonic case
of section \ref{sigma}. In other words, the selfdual supersymmetric
$\sigma$-model we have defined in this section is indeed a supersymmetric
extension of the bosonic one.

\subsection{Free scalar superfield and T-duality}

Going back to our trivial example, an abelian algebra with two
generators $h_1$ and $h_2$, we have
\be
\cV = e^\Phi
\ee
with\footnote{There is a $-i$ factor in the fields definition in order 
to simplify formulae below. In Lorentzian signature, there would not be such
$i$ factors everywhere.}
\be
\Phi = -i \Phi_1 h_1 + \Phi_2 h_2 \rlap{\ ,}
\ee
which gives a field strength
\be
\cG = -i \cG_1 h_1 + \cG_2 h_2 
= \G(\begin{array}{c} -i D \Phi_1 h_1 + D \Phi_2 h_2 \\ 
-i \bD \Phi_1 h_1 + \bD \Phi_2 h_2 \end{array}\D)\rlap{\ .}
\ee

The Bianchi identities
\bea
D \bG_1 + \bD G_1 &=& 0 \nonumber\\
D \bG_2 + \bD G_2 &=& 0
\eea
are trivial as the algebra is abelian and 
are simply $[D,\bD]=0$ applied to $\Phi_1$ and $\Phi_2$.

With $\cS$ exchanging $h_1$ and $h_2$ as above, the selfduality equation is
\bea
G_1 &=& G_2 \nonumber\\
- \bG_1 &=& \bG_2 \rlap{\ .}
\eea

Thus, the Bianchi identity for $\cG_2$ becomes the equation of motion of
$\cG_1$:
\be
D \bG_1 - \bD G_1 = 0 \rlap{\ ,}
\ee
that is
\be
D \bD \Phi_1 = 0 \rlap{\ ,}
\ee
which describes the motion of a free massless scalar superfield.
If we write in components
\be
\Phi_1 = \phi_1 + \ath \psi_1 + \bth \widetilde{\psi}_1 + \ath\bth F_1
\rlap{\ ,}
\ee
we get the usual equations
\bea
\p_z \p_\bz \phi_1 &=& 0 \\
\p_\bz \psi_1 &=& 0 \\
\p_z \widetilde{\psi}_1 &=& 0 \\
F_1 &=& 0 \rlap{\ .}
\eea

To emphasize the physical meaning of selfduality, let us write it in term
of $\Phi_i$'s:
\bea
D\Phi_1 &=& D\Phi_2 \nonumber \\
- \bD\Phi_1 &=& \bD\Phi_2 \rlap{\ .}
\label{T-dual}
\eea
With see that we can write $\Phi_1$ and $\Phi_2$ as
\bea
\Phi_1 &=& \Phi_{c} + \Phi_{ac} \nonumber\\
\Phi_2 &=& \Phi_{c} - \Phi_{ac}
\eea
in terms of a chiral superfield $\Phi_c$ and an antichiral $\Phi_{ac}$. It
makes apparent that the duality involved here is simply \emph{T-duality} of
a one-dimensional target-space in the worldsheet perspective. It means that
the equations of motion follow from the existence of T-duality written as in
(\ref{T-dual}). With more flat target-space dimensions, equations of motion
are equivalent to the possibility of T-dualize all dimensions.

\subsection{More supersymmetries}

Generalization to a larger number of supercharges is straightforward. 
As we consider massless scalars, the $(N,\widetilde{N})$ supersymmetry
algebra is
\bea
\G[Q^I,Q^{I'}\D] &=& -\delta^{II'} \p_z \nonumber\\
\G[\overline{Q}^J,\overline{Q}^{J'}\D] &=& -\delta^{JJ'} \p_\bz \nonumber\\
\G[Q^I,\overline{Q}^J\D] &=& 0 \rlap{\ .}
\eea
In superspace, we have odd variables $\ath^I$ and $\bth^J$, $I = 1
\ldots N$, $J = 1 \ldots \widetilde{N}$. The covariant derivatives are
\bea
D^I&=&\p_{\ath^I} + \ath^I \p_z \nonumber\\
\bD^J &=&\p_{\bth^J} + \bth^J \p_\bz \rlap{\ .}
\eea

The field strength $\cG$ has $N$ left and $\widetilde{N}$ right components
\bea
G^I &=& D^I\cV \, \cV^{-1} \nonumber\\
\bG^J &=& \bD^J\cV \, \cV^{-1}
\eea 
with Bianchi identities
\be
\bD^J G^I + D^I \bG^J = \G[ G^I,\bG^J \D] \rlap{\ .}
\ee

The Hodge star $*$ acts for all values of indices as 
\bea
G^I &\longrightarrow& -i \, G^I \nonumber\\
\bG^J &\longrightarrow& i \, \bG^J \rlap{\ ,}
\eea
such that the selfduality equation $*\,\cS\cG=\cG$ reads
\bea
-i\cS G^I &=& G^I \nonumber\\
i\cS \bG^J &=& \bG^J \rlap{\ .}
\eea
It is essential that the sign is the same for all values of the indices:
one can change a global sign in $\cS$ but left and right components must
take well defined opposite signs.

Equations of motion are obtained by introducing this selfduality equation 
into the Bianchi identities:
\be
-i \bD^J \! \cS G^I +i D^I \! \cS\bG^J = \G[ G^I,\bG^J \D] \rlap{\ .}
\ee

If we truncate a $(N,\widetilde{N})$ superfield $\Phi$ to its lower component
$\phi$, all of these equations of motion give back the equation of the
bosonic selfdual $\sigma$-model, exactly as in section \ref{sigma-11}.

\subsection{$(N,0)$ supersymmetry}

The $(N,0)$-superconformal case is slightly different: one has covariant
$D^I$ in the left sector and the usual derivative $\bp$ in the right one.

The field strength $\cG$ has now $N$ fermionic components 
on the left and one bosonic component on the right:
\bea
G^I &=& D^I\cV \, \cV^{-1} \nonumber\\
G_\bz &=& \p_\bz \cV \, \cV^{-1} \rlap{\ .}
\eea
The Bianchi identities now involve commutators:
\be
D^I G_\bz -\p_\bz G^I = \G[ G^I , G_\bz \D] \rlap{\ .} 
\ee

The Hodge star $*$ obviously acts on $\cG$ as
\bea
G^I &\longrightarrow& -i \, G^I \nonumber\\
G_\bz &\longrightarrow& i \, G_\bz \rlap{\ ,}
\eea
and the selfduality equation $*\,\cS\cG=\cG$ reads
\bea
\cS G^I &=& G^I \nonumber\\
-\cS G_\bz &=& G_\bz \rlap{\ ,}
\eea
giving the equations of motion
\be
i D^I\! \cS G_\bz +i\p_\bz \cS G^I = \G[ G^I , G_\bz \D] \rlap{\ .}
\ee

If we truncate the $\mathfrak{G}$-valued superfield $\cV=e^\Phi$ to 
its lowest component $e^\phi$, we get\footnote{Remember $\cG^{(0)} = 
G_z^{(0)} \ud z + G_\bz^{(0)} \ud \bz$ is the bosonic field strength 
derived from $\phi$.}
\bea
G^I &=& D^I e^\phi \, e^{-\phi}
= \ath^I \p_z \phi + \frac{1}{2!} [ \phi, \ath^I \p_z \phi]
+ \frac{1}{3!} [\phi, [\phi, \ath^I \p_z \phi]] + \ldots = \ath^I \,
G_z^{(0)}
\nonumber\\
G_\bz &=& \p_\bz e^\phi \, e^{-\phi}
= \p_\bz \phi + \frac{1}{2!} [ \phi, \p_\bz \phi]
+ \frac{1}{3!} [\phi, [\phi, \p_\bz \phi]] + \ldots = G_\bz^{(0)}
\eea
and the Bianchi identity can be written as
\be
\ath^I \p_z G_\bz^{(0)} - \ath^I \p_\bz G_z^{(0)} 
= \ath^I \G[ G_z^{(0)} , G_\bz^{(0)} \D]
\rlap{\ ,}
\ee
which is the Bianchi identity of the bosonic $\sigma$-model.
The Hodge star gives naturally the usual one on the complex plane, so that
we recover the bosonic selfdual $\sigma$-model with this truncation
of the $(N,0)$ theory. It is easy to see that all superconformal
models defined here contained those with less supersymmetry as truncations,
as one expects. The only condition is, trivially, to keep the same algebra
$\mathfrak{g}$.

\section{Selfduality in symmetric spaces}
\label{self-sym}

\subsection{Symmetric spaces}

There is a class of models which are of great interest, when $\cV$ lives in 
the symmetric space $\mathfrak{G}/\mathfrak{H}$. $\mathfrak{H}$ is a
maximal subgroup of $\mathfrak{G}$, defined as the set of fixed points of an
involution $\tau$ acting on $\mathfrak{G}$. It induces an involution on the
tangent algebra $\mathfrak{g}$, which we still denote by $\tau$, and which
gives a decomposition
\be
\mathfrak{g} = \mathfrak{h} \oplus \mathfrak{h}^\perp \rlap{\ ,}
\ee
where $\mathfrak{h}$ and $\mathfrak{h}^\perp$ are sets of respectively
invariant and antiinvariant elements of $\mathfrak{g}$ under the action of 
$\tau$. We have the maximality property
\bea
\G[ \mathfrak{h} , \mathfrak{h} \D] &\subset & \mathfrak{h} \nonumber\\
\G[ \mathfrak{h} , \mathfrak{h}^\perp \D] &\subset & \mathfrak{h}^\perp
\nonumber\\
\G[ \mathfrak{h}^\perp , \mathfrak{h}^\perp \D] &\subset & \mathfrak{h}
\rlap{\ .}
\label{sym}
\eea

A special case of physical interest occurs when $\mathfrak{H}$ is the
maximal compact subgroup of $\mathfrak{G}$, with $\tau$ the Cartan involution.
Note also that any group Lie $\mathfrak{G}$ can be seen as the
symmetric space $(\mathfrak{G} \times \mathfrak{G})/\Delta \mathfrak{G}$,
where $\Delta \mathfrak{G}$ is $\mathfrak{G}$ acting diagonally on
$\mathfrak{G} \times \mathfrak{G}$:
\be
g.(g_1,g_2) = (gg_1,gg_2)
\rlap{\ .}
\ee
In this case, one recovers principal chiral models, here in a superconformal
version.

Although we could work with an arbitrary number of supercharges,
we work now with $(1,1)$ supersymmetry. Generalization to other
cases follows easily, as we have seen in last section for a simpler model. 

So let $\cV$ be a superfield with values in $\mathfrak{G}/\mathfrak{H}$,
parametrized by superfields $\phi^i$. We define the field strength $\cG =
\mathcal{X} + \mathcal{Y}$
as
\bea
G &=& D \cV \, \cV^{-1} = X + Y \nonumber\\
\bG &=& \bD \cV \, \cV^{-1} = \bX + \bY \rlap{\ ,}
\eea
where $X$ and $\bX$ have values in $\mathfrak{h}$, $Y$ and $\bY$ 
in $\mathfrak{h}^\perp$.

Because of (\ref{sym}), Bianchi identities decomposes on $\mathfrak{h}$ and
$\mathfrak{h}^\perp$ as
\bea
\bD X + D \bX - \G[ X, \bX \D] &=& \G[ Y , \bY \D] \nonumber\\
\bcD Y + \cD \bY &=& 0 \rlap{\ ,}
\label{sigma-bianchi}
\eea
where $\cD$ and $\bcD$ are covariant derivatives with respect to
$\mathfrak{H}$:
\bea
\cD \bY &=& D \bY - \G[ X , \bY \D] \nonumber\\ 
\bcD Y &=& D Y - \G[ \bX , Y \D] \rlap{\ .}
\eea

$\cG$ is invariant under global $\mathfrak{G}$ transformations acting on the
right and covariant under local $\mathfrak{H}$ on the left:
\be
\cV(z,{\bar z}, \ath,\bth) \ \longrightarrow \ \Xi(z,{\bar z},\ath,\bth) 
\, \cV(z,{\bar z}, \ath,\bth) \, \Lambda \rlap{\ .}
\ee
Invariance under a global $\Lambda \in \mathfrak{G}$ is the same as above. 
Gauge action under $\Xi(z,{\bar z},\ath,\bth) \in \mathfrak{H}$ acts on $G$ as
\be
D\cV\,\cV^{-1} \ \longrightarrow \ \Xi\, D\cV\,\cV^{-1} \,\Xi^{-1} +
D\Xi\,\Xi^{-1} \rlap{\ .}
\ee
From (\ref{sym}) and $\Xi \in \mathfrak{H}$ we get in terms of $X$ and $Y$
\bea
X &\longrightarrow& \Xi\, X \,\Xi^{-1} + D\Xi\,\Xi^{-1} \nonumber \\
Y &\longrightarrow& \Xi\, Y \,\Xi^{-1} \rlap{\ .}
\eea
For the right components, we get similarly
\bea
\bX &\longrightarrow& \Xi\, \bX \,\Xi^{-1} + \bD\Xi\,\Xi^{-1} \nonumber \\
\bY &\longrightarrow& \Xi\, \bY \,\Xi^{-1} \rlap{\ .}
\eea
We check that $\mathcal{X} = \G(\begin{array}{c} X \\ \bX \end{array}\D)$
transforms indeed as a gauge field.

This gauge invariance allow us to parametrized $\mathfrak{G}/\mathfrak{H}$
with a fixed set of representatives in $\mathfrak{G}$. A common choice for
$\mathfrak{G}$ a simple group and $\mathfrak{H}$ its maximal compact
subgroup is the Borel gauge: $\cV$ is taken in the Borel subgroup
$\mathfrak{B}$ of $\mathfrak{G}$.  The Borel subalgebra $\mathfrak{b}$ 
is spanned by Cartan elements and
generators associated to positive roots. We will use such a gauge choice
when we deal with the $SL(2)/SO(2)$ example. Global $\mathfrak{G}$ 
transformations do not always preserve the gauge. In the Borel gauge, only
elements of the Borel subgroup leave the gauge invariant. Other
transformations must be compensated by $\mathfrak{H}$ transformations
(Iwasawa decomposition ensures it is always possible), 
such that
\be
\Xi(\cV,\Lambda) \, \cV \, \Lambda \in \mathfrak{B} \rlap{\ .}
\ee

\subsection{Superconformal selfduality}

We want now to get equations of motion by imposing a selfduality constraint.
We need an $\mathfrak{H}$-invariant pseudo-involution\footnote{It has square
$-1$ in the Euclidean setting, but, as before, it would be a 
real involution with square identity in the Lorentzian case: $\cS^2*^2 =
1$.} $\cS$ acting on $\mathfrak{g}$:
\be
\cS = \Xi \, \cS \, \Xi^{-1} 
\ee
for any $\Xi \in \mathfrak{H}$.

It allows us to write a gauge-invariant selfduality equation
\be
* \cS \mathcal{Y} = \mathcal{Y} \rlap{\ ,}
\label{sym-dual}
\ee
\ie
\bea
-i \cS Y &=& Y \nonumber\\
i \cS \bY &=& \bY \rlap{\ .}
\eea

With the second Bianchi identity of (\ref{sigma-bianchi}) applied to the
field $\cS \mathcal{Y}$, we get the equation of motion
\be
\bcD Y - \cD \bY = 0 \rlap{\ .}
\label{sym-motion}
\ee

This model has several symmetries: superconformal symmetry on the
worldsheet, global $\mathfrak{G}$ and local $\mathfrak{H}$ in the target
space $\mathcal{M} = \mathfrak{G}/\mathfrak{H}$.

\subsection{$\sigma$-model action}
\label{sigma-def}

Though our aim is to describe models where the equations of motion come from
a selfduality constraint, it is worth remarking that 
the equation of motion (\ref{sym-motion}) extremizes the superconformal
$\sigma$-model action \cite{dv-f,w}
\be
S = \frac{1}{2} \int \ud z^2 \ud\theta^2 \ \mathrm{Tr}\G( \mathcal{Y} \w 
* \mathcal{Y} \D) \rlap{\ ,}
\label{sigma-action}
\ee
with the superspace Hodge star $*$ defined above. The trace is computed
using an $\mathfrak{h}$ invariant bilinear tensor $\eta_{ab}$ on 
$\mathfrak{g}$. (In some cases, there are several inequivalent invariant 
tensors $\eta$.) The wedge product is defined as follows:
\be
\mathcal{Y} \w \mathcal{Y'} = Y \bY' - \bY Y'
\rlap{\ .}
\ee
The equation of motion is indeed
\be
\bcD Y - \cD \bY = 0
\rlap{\ .}
\ee

We have showed here that selfdual sets of fields are special solutions of
these models. We will prove in section \ref{loop} that all solutions to this
equation of motion can be seen as selfdual configurations, if one introduces
a infinite set of fields.

\subsection{Example: $SL(2)/SO(2)$}

We take $\mathfrak{G}=SL(2)$, and $\mathfrak{H}$ its maximal compact
subgroup $SO(2)$, characterized by the Cartan involution
\be
O \ \longrightarrow \ {}^t O^{-1} \rlap{\ ,}
\ee
which gives on the tangent algebra
\be
\tau(M) = -M \rlap{\ .}
\ee
We take
\be
U = \G(\begin{array}{cc} 0 & 1 \\
      -1 & 0 \end{array} \D)
\ee
as basis element of $\mathfrak{h} = \mathfrak{so}(2)$
and
\bea
V^1 &=& \G(\begin{array}{cc} 0 & 1 \\
      1 & 0 \end{array} \D) \nonumber\\
V^2 &=& \G(\begin{array}{cc} 1 & 0 \\
      0 & -1 \end{array} \D)
\eea
as basis of $\mathfrak{h}^\perp$.

We parametrize $SL(2)/SO(2)$ elements in the Borel gauge of upper triangular
matrices as
\be
\cV = \G(\begin{array}{cc} e^\phi & N e^\phi \\
      0 & e^{-\phi} \end{array} \D)
\ee
with real superfields $N$ and $\phi$.

The field strength is
\bea
G = D\cV \, \cV^{-1} &=& \G(\begin{array}{cc} D\phi & e^{2\phi} DN\\
                       0 & -D \phi \end{array} \D) \nonumber\\
\bG = \bD\cV \, \cV^{-1} &=& \G(\begin{array}{cc} \bD\phi & e^{2\phi} \bD N\\
                       0 & -\bD \phi \end{array} \D)
\rlap{\ .}
\eea
It decomposes on $\mathfrak{h} \oplus \mathfrak{h}^\perp$ as
\bea
G &=& \frac{1}{2} e^{2\phi} DN \, U + \frac{1}{2} e^{2\phi} DN \, V^1
+ D\phi \, V^2 \nonumber\\
\bG &=& \frac{1}{2} e^{2\phi} \bD N \, U + \frac{1}{2} e^{2\phi} \bD N \, V^1
+ \bD\phi \, V^2 \rlap{\ .}
\eea

A selfduality constraint can be imposed if we have some operator $\cS$
acting on $\mathfrak{h}^\perp$ with
$\cS^2=-1$ and which commutes with $\mathfrak{so}(2)$. There is a unique
solution, up to a global minus sign:
\bea
\cS V^1 &=& -V^2 \nonumber\\
\cS V^2 &=& V^1 \rlap{\ .}
\eea

The selfduality equation $* \cS \mathcal{Y} = \mathcal{Y}$ reads
\bea
-i \cS \G( \frac{1}{2} e^{2\phi} DN \, V^1 + D\phi \, V^2 \D) 
&=& \frac{1}{2} e^{2\phi} DN \, V^1 + D\phi \, V^2 \nonumber\\
i \cS \G( \frac{1}{2} e^{2\phi} \bD N \, V^1 + \bD\phi \, V^2 \D)
&=& \frac{1}{2} e^{2\phi} \bD N \, V^1 + \bD\phi \, V^2
\eea
\ie
\bea
\frac{i}{2} e^{2\phi} DN &=& D\phi \nonumber\\
-\frac{i}{2} e^{2\phi} \bD N &=& \bD\phi \rlap{\ .}
\eea
This selfduality constraint is easily solved in terms of chiral and
antichiral superfields $f$ and $\tilde{f}$:
\bea
e^{-2\phi} &=& \frac{1}{2} \G( f(z,\ath) + \tilde{f}({\bar z},\bth) \D)
\nonumber\\
N &=& -\frac{1}{2i} \G( f(z,\ath) - \tilde{f}({\bar z},\bth) \D)
\rlap{\ .}
\eea
The reality of $\phi$ and $N$ imposes $\tilde{f} = \bar{f}$ and finally the
selfdual configurations are expressed in term of a single chiral superfield
$f$ as
\bea
e^\phi &=& \frac{1}{\sqrt{\Re(f)}} \nonumber\\
N &=& -\Im(f) \rlap{\ .}
\eea
As we have seen above, these selfdual configurations are in particular 
solutions to the equations of motion of the $\sigma$-model action 
(\ref{sigma-action}).

\section{Loop group symmetry}
\label{loop}

\subsection{Loop group $\sigma$-model}

We show now how the symmetric space $\mathfrak{G}/\mathfrak{H}$
$\sigma$-model can be seen as an infinite-dimensional
$\widetilde{\mathfrak{G}}/\widetilde{\mathfrak{H}}$ $\sigma$-model with a
self-duality contraint, where $\widetilde{\mathfrak{G}}$ is the loop group
extension of $\mathfrak{G}$ and $\widetilde{\mathfrak{H}}$ a maximal
subgroup. By construction, it will be endorsed with global
$\widetilde{\mathfrak{G}}$ symmetry and local $\widetilde{\mathfrak{H}}$
gauge symmetry. We still work with $(1,1)$ supersymmetry, but it is
straightforward to do it with a different or vanishing number of
supercharges.

The loop extension $\widetilde{\mathfrak{G}}$ of $\mathfrak{G}$ is
the infinite-dimensional Lie group which has tangent Lie algebra
\be
\widetilde{\mathfrak{g}} = \RR\G[t,\frac{1}{t}\D] \otimes \mathfrak{G}
\rlap{\ ,}
\ee
the loop algebra extension of $\mathfrak{g}$.
In practice, we deal with elements of this group and this algebra as 
functions of the spectral parameter $t$ into respectively 
$\mathfrak{G}$ and $\mathfrak{g}$.

Let $\mathfrak{h}$ be a maximal subalgebra of $\mathfrak{g}$, defined by an
involution $\tilde{\tau}$. $\tau$ is extended on $\widetilde{\mathfrak{g}}$
by\footnote{This is for Euclidean signature. For a Lorentzian worldsheet, the
minus sign would not be there.}
\be
\tilde{\tau}(\mathcal{A}(t)) = \tau\G( \mathcal{A}\G(-\frac{1}{t}\D) \D) 
\rlap{\ ,}
\ee
with $\mathcal{A}(t) \in \widetilde{\mathfrak{g}}$ \cite{brei-mai,ju1,ju-ni}.
With an angle parameter $\alpha$, $t=\tan\G(\frac{\alpha}{2}\D)$, this
involution exchanges $\alpha$ and $\alpha+\pi$:
\be
\tilde{\tau}(\mathcal{A}(\alpha)) = \tau(\mathcal{A}(\alpha+\pi))
\rlap{\ .}
\ee
We denote by $\widetilde{\mathfrak{h}}$ the subalgebra of fixed points of
$\widetilde{\mathfrak{g}}$ by $\tilde{\tau}$. It is not the loop
extension of $\widetilde{\mathfrak{h}}$.
The involution is defined in the same way on the group
$\widetilde{\mathfrak{G}}$ itself, with the subgroup of fixed points
$\widetilde{\mathfrak{H}}$.

The component of degree zero in a $t$ expansion of
$\widetilde{\mathfrak{G}}$ is the original group $\mathfrak{G}$. As
$\tilde{\tau}$ reduces to $\tau$ on that component, the $t=0$ part of
$\widetilde{\mathfrak{H}}$ is $\mathfrak{H}$. Of course, the same is true
for tangent algebras. Note that $\tfH$ is not the loop extension of
$\mathfrak{H}$.

We take a field $\tcV$ in the coset symmetric space
$\widetilde{\mathfrak{G}}/\widetilde{\mathfrak{H}}$. Due to the gauge
freedom, we choose representatives in a "Borel gauge": $\tcV(t)$ must be
analytic inside a unit disc around zero \cite{brei-mai,ni2,ju-ni,kns}.
(Note that the choice of this point
is arbitrary. It will allow to recover physical quantities as the $t=0$
value of fields.) The existence and uniqueness of such a gauge is a
Riemann-Hilbert problem: if we have $\tcV \in \widetilde{\mathfrak{G}}$, we
can decompose $\tcV \, \tilde{\tau}\!\G(\tcV^{-1}\D)$ as $\tcV_+ \,
\tilde{\tau}\!\G(\tcV_-^{-1}\D)$, with $\tcV_+$ analytic inside the unit disc
and $\tcV_-$ analytic outside. ($t$ lives on the Riemann sphere.) We are
left with gauge freedom at $t=0$, that we fix as in the
$\mathfrak{G}/\mathfrak{H}$ $\sigma$-model: we set $\tcV(0)$ to be in a
given Borel subgroup of $\mathfrak{G}$.

The field strength
is\footnote{$D$ and $\bD$ derivatives acts on superspace variables
$z$, $\bar z$, $\ath$ and $\bth$, not on the spectral parameter $t$.}
\be
\tcG = \G( \begin{array}{c} \tG \\ \btG \end{array} \D)
= \G( \begin{array}{c} D\tcV \, \tcV^{-1} \\ \bD\tcV \, \tcV^{-1} \end{array}
\D)
\rlap{\ .}
\ee
A Bianchi identity follows naturally from the definition:
\be
D\btG + \bD\tG = \G[ \tG , \btG \D] \rlap{\ .}
\label{loop-bianchi}
\ee

If $\tcV$ is in the Borel gauge defined above, $\tcG$ has an expansion
\be
\tcG = \mathcal{X} + \mathcal{Y} + \sum_{n > 0} \cA_n t^n \rlap{\ ,}
\ee
with
\bea
\mathcal{X} & \in & \mathfrak{h} \nonumber\\
\mathcal{Y} & \in & \mathfrak{h}^\perp \nonumber\\
\mathcal{\cA}_n & \in & \mathfrak{g} \rlap{\ .}
\eea
The decomposition $\tcG = \tcX + \tcY$ on $\tfh \oplus \tfh^\perp$ is
\bea
\tcX &=& \mathcal{X} + \frac{1}{2} \sum_{n > 0} \cA_n t^n
+ \frac{1}{2} \sum_{n > 0} \tau(\cA_n) \G(-\frac{1}{t}\D)^n \nonumber\\
\tcY &=& \mathcal{Y} + \frac{1}{2} \sum_{n > 0} \cA_n t^n
- \frac{1}{2} \sum_{n > 0} \tau(\cA_n) \G(-\frac{1}{t}\D)^n
\rlap{\ .}
\eea

\subsection{Selfduality constraint}

In order to impose a selfduality constraint, we define the operator $\cS$ in
the following way:
\be
\cS : \cA(t) \ \longrightarrow \ -\, t\, \tilde{\tau}(\mathcal{A}(t)) 
\rlap{\ .}
\ee
It is not hard to check that $\cS$ is a pseudo-involution\footnote{In a
Lorentzian signature, it would square to $+1$.}:
\be
\cS^2 = -1 \rlap{\ ,}
\ee
such that the full duality operator $*\cS$ is a real involution:
\be
(* \cS)^2 = 1 \rlap{\ .}
\ee

As in the finite-dimensional case, we impose the selfduality constraint
\be
*\cS\tcY = \tcY \rlap{\ ,}
\label{loop-self}
\ee
where $*$ is here the superspace Hodge star defined in section
\ref{sigma-11},
but in a purely bosonic case would be the usual Hodge operator.

For an element $\widetilde{\cY}$ of $\tfh^\perp$, which by definition
satisfies
\be
\tilde{\tau}(\widetilde{\cY}) = - \widetilde{\cY}
\rlap{\ ,}
\ee
we have
\be
\cS \tcY(t) = t \tcY(t)
\rlap{\ .} 
\ee
The selfduality constraint (\ref{loop-self}) is thus equivalent to
\be
*\cT\tcY = \tcY \rlap{\ ,}
\label{t-constraint}
\ee
where $\cT$ is the operator shifting elements of $\tfg$ by one degree in
$t$:
\be
\cT : \cA(t) \ \longrightarrow \ t \,\cA(t)
\rlap{\ .}
\ee

In components of the $t$-expansion of $\tcY$, the selfduality equation
gives
\bea
\cA_{2p} &=& 2(-1)^p \, \mathcal{Y} \nonumber\\
\cA_{2p+1} &=& 2(-1)^p *\!\mathcal{Y} \rlap{\ .}
\eea

If we sum up all components, we get for $\tcG$
\be
\tcG = \cX + \frac{1-t^2}{1+t^2} \, \cY + \frac{2t}{1+t^2} *\!\cY
\rlap{\ ,}
\ee
which is well known as the Lax pair equation\footnote{In Lorentzian
signature,
it would be
$\tcG = \cX + \frac{1+t^2}{1-t^2} \, \cY + \frac{2t}{1-t^2} *\!\cY$.}
 associated to the
$\mathfrak{G}/\mathfrak{H}$ $\sigma$-model described in section
\ref{sigma-def} \cite{fgs,hruby}.
In term of the angle variable $\alpha$, it reads
\be
\widetilde{\cG} = \cX + \cos(\alpha)  \, \cY + \sin(\alpha)  *\!\cY
\rlap{\ .}
\ee
In superconformal coordinates, with covariant derivatives
\bea
\cD \tcV &=& D \tcV - X \tcV \nonumber\\
\bcD \tcV &=& \bD \tcV - \bX \tcV \rlap{\ ,}
\eea
we have
\bea
\cD \tcV \, \tcV^{-1} &=& \frac{i+t}{i-t} \ Y \nonumber\\
\bcD \tcV \, \tcV^{-1} &=& \frac{i-t}{i+t} \ \bY \rlap{\ .}
\label{lax}
\eea
It is a linear system which allows to get $\tcV$ by integration,
at least formally.

The Bianchi identity and the equation of motion of the original field
\be
\cV = \tcV(0)
\ee
follow from the compatibility condition, or Maurer-Cartan equation, of this
linear system. The Bianchi identity of $\cV$ comes simply from the $t=0$
(\ie $\alpha=0$) part
of (\ref{loop-bianchi}). Indeed, the truncation of (\ref{lax}) to $t=0$ is
the defining equation of $Y$ and $\bY$ in term of $\cV$, and we find back
Bianchi identities (\ref{sigma-bianchi}).

The equation of motion comes from the Bianchi identity of $\tcG$ at 
$\theta=\frac{\pi}{2}$ ($t=1$):
\be
\cD \bY - \bcD Y = 0 \rlap{\ .}
\ee

\subsection{Symmetries}

By construction, this model has global $\tfG$ symmetry and local $\tfH$
gauge symmetry, because $\cS$ commutes with $\tfH$. Indeed, if we act on
$\tcV$ with a global $\widetilde{\Lambda} \in \tfG$, we may have to
compensate by an $\tfH$ gauge transformation $\widetilde{\Xi}$ to remain 
in the Borel gauge:
\be
\tcV \ \longrightarrow \ \widetilde{\Xi} \, \tcV \, \widetilde{\Lambda} \rlap{\ .}
\ee
Under such a transformation, $\tcY$ becomes
\be
\tcY \ \longrightarrow \ \widetilde{\Xi} \, \tcY \, \widetilde{\Xi}^{-1}
\rlap{\ .}
\ee
The  constraint (\ref{t-constraint}), equivalent to the selfduality
constraint (\ref{loop-self}), is invariant provided
\be
\widetilde{\Xi} \, \cT \, \widetilde{\Xi}^{-1} = \cT \rlap{\ ,}
\ee
which is indeed verified.
Under the compensating $\tfH$ gauge transformation, $\tcX$ varies as a gauge
field:
\bea
\widetilde{X} &\longrightarrow& \widetilde{\Xi}\, \widetilde{X}
\,\widetilde{\Xi}^{-1} + D\widetilde{\Xi}\,\widetilde{\Xi}^{-1} \nonumber \\
\widetilde{\bX} &\longrightarrow& \widetilde{\Xi}\, \widetilde{\bX}
\,\widetilde{\Xi}^{-1} + \bD\widetilde{\Xi}\,\widetilde{\Xi}^{-1}
\rlap{\ .}
\eea
One recovers transformation of physical field strength 
$\mathcal{X}$ and $\mathcal{Y}$ by taking the $t=0$ component of
transfomration laws of $\tcX$ and $\tcY$.

$\tfG$ transformations $g\,t^n$ with $n>0$ have no effect on the physical
field $\cV$: they do not require gauge compensation and modify only 
integration constants of higher levels of $\tcV$ \cite{ju-ni}. For $n=0$, we
recover the usual symmetry of the $\mathfrak{G}/\mathfrak{H}$ $\sigma$-model. 
For $n<0$, physical fields are modified by gauge compensation
$\widetilde{\Xi}$. In fact, these non-local additional symmetries can be
used to generate new (classical) solutions of the model \cite{dolan,ei-fo,wu}. 
The fact that some of the transformations are hidden whereas others are
visible through the gauge compensation is due to the fixing of the gauge
that we use to extract physical fields. If we restore gauge freedom, we
cannot get rid of the full group of symmetry $\tfG$. It should be noted
that, if the rigid symmetry group is the same as in the bosonic case, in
supersymmetric models the gauge symmetry is enlarged: gauge transformations 
depend on both bosonic and fermionic coordinates.

\subsection*{Aknowledgments}

We thank B.~Julia for discussions on related topics, and also A.~Keurentjes
for comments. This work is
supported in part by the "Actions de Recherche Concertées" of
the "Direction de la Recherche Scientifique - Communauté
Française de Belgique", by a "Pôle d'Attraction
Interuniversitaire" (Belgium) and by IISN-Belgium (convention
4.4505.86). Support from  the European Commission RTN programme
HPRN-CT-00131, in which we are associated to K.~U.~Leuven, is also 
acknowledged.

\end{document}